\documentclass[preprint,aps,showpacs,nofootinbib,preprintnumbers,amsmath,amssymb]{revtex4-1}
\usepackage{epsfig}
\usepackage{caption}
\usepackage{subfigure}
\usepackage{dcolumn}
\usepackage{bm}
\usepackage[usenames ,dvipsnames]{xcolor}
\usepackage{slashed}
\usepackage{graphicx,color}

\begin{document}
\title{SU(3) symmetry breaking in charmed baryon decays}

\author{C.Q. Geng$^{1,2}$, Y.K. Hsiao$^{1,2}$, Chia-Wei Liu$^{2}$ and Tien-Hsueh Tsai$^{2}$}
\affiliation{
$^{1}$School of Physics and Information Engineering, Shanxi Normal University, Linfen 041004, China\\
$^{2}$Department of Physics, National Tsing Hua University, Hsinchu, Taiwan 300
}\date{\today}

\begin{abstract}
We explore the breaking effects of the $SU(3)$ flavor symmetry in the singly Cabibbo-suppressed anti-triplet charmed baryon decays of ${\bf B}_c\to {\bf B}_n M$, with ${\bf B}_c=(\Xi_c^0,\Xi_c^+,\Lambda_c^+)$ and ${\bf B}_n(M)$  the baryon (pseudo-scalar) octets. We find that these breaking effects can be used to account  for the experimental data on the decay branching ratios of ${\cal B}(\Lambda_c^+\to \Sigma^{0} K^{+},\Lambda^{0} K^{+})$ and $R'_{K/\pi}$=${\cal B}(\Xi^0_c \to \Xi^- K^+)$/${\cal B}(\Xi^0_c \to \Xi^- \pi^+)$.
In addition, we obtain that ${\cal B}(\Xi_{c}^{0}  \to  \Xi^{-} K^{+},\Sigma^{-} \pi^{+})=(4.6 \pm 1.7,12.8 \pm 3.1)\times 10^{-4}$, ${\cal B}(\Xi_c^0\to pK^-,\Sigma^+\pi^-)=(3.0 \pm 1.0, 5.2 \pm 1.6)\times 10^{-4}$ and ${\cal B}(\Xi_c^+\to \Sigma^{0(+)} \pi^{+(0)})=(10.3 \pm 1.7)\times 10^{-4}$, which all receive significant contributions from the breaking effects, and can be tested by the BESIII and LHCb experiments.
\end{abstract}

\maketitle
\section{Introduction}
It is known that
the theoretical approach based on the factorization and
quantum chromodynamics (QCD)
barely explains the charmed hadron decays~\cite{Khodjamirian:2017zdu}.
This is due to the fact that the mass of the charm quark, $m_c\simeq 1.5$~GeV, is not as heavy as that of the bottom one,
$m_b\simeq 4.8$~GeV, resulting in an undetermined correction to the heavy quark expansion, such that
the alternative models have to take place for this correction~\cite{Cheng:1991sn,Cheng:1993gf,
Zenczykowski:1993hw,Fayyazuddin:1996iy,Dhir:2015tja,Cheng:2018hwl}.
On the other hand,
the $SU(3)$ flavor ($SU(3)_f$) symmetry
that works in the $b$-hadron decays~\cite{He:2000ys,Fu:2003fy,Hsiao:2015iiu,He:2015fwa,He:2015fsa}
can be well applied to $D\to MM$ and
${\bf B}_c\to {\bf B}_n M$~\cite{Cheng:2012xb,Pirtskhalava:2011va,Grossman:2012ry,Savage:1989qr,
Savage:1991wu,h_term,Lu:2016ogy,Geng:2017esc,Geng:2018plk,Wang:2017gxe,
Wang:2017azm,Geng:2017mxn},
where ${\bf B}_c=(\Xi_c^{0},-\Xi_c^{+},\Lambda_c^+)$ are
the lowest-lying anti-triplet charmed baryon states, while
${\bf B}_n$ and $M$ represent baryon and pseduscalar meson states, respectively.
Particularly, the $SU(3)_f$ symmetry has been extended to investigate
the singly charmed baryon sextet states as well as the doubly and triply
charmed baryon ones~\cite{Wang:2017azm,Geng:2017mxn}.
For  $D\to MM$ decays, the measurements produce~\cite{pdg}
\begin{eqnarray}\label{data1}
{\cal R}_{D^0(K/\pi)}&\equiv  &\frac{{\cal B}(D^0\to K^+K^-)}{{\cal B}(D^0\to \pi^+\pi^-)} = 2.82\pm 0.07\,,
\nonumber\\
{\cal B}_{D^0(2K_s^0)}&\equiv&{\cal B}(D^0\to K_s^0K_s^0)=(1.70\pm 0.12)\times 10^{-4}\,,
\end{eqnarray}
in comparison with $({\cal R}_{D^0(K/\pi)},{\cal B}_{D^0(2K_s^0)})\simeq (1,0)$
given by the theoretical calculations  based on the $SU(3)_f$ symmetry.
The disagreements between the theory and experiment imply that
 the   breaking effects of the $SU(3)_f$ symmetry cannot be ignored
 in the singly Cabibbo-suppressed (SCS) processes.
 We note that,
 in the literature,
the  $SU(3)_f$  breaking effects were used to  relate ${\cal R}_{K/\pi}$ to the possible large difference of the $CP$ violating asymmetries
of $\Delta {\cal A}_{CP}\equiv {\cal A}_{CP}(D^0\to K^+K^-)
-{\cal A}_{CP}(D^0\to \pi^+\pi^-)$~\cite{Brod:2011re,Pirtskhalava:2011va,Brod:2012ud},
which is recently measured to be $(-0.10\pm 0.08\pm 0.03)\%$ by LHCb~\cite{Aaij:2016cfh}.

For the two-body ${\bf B}_c\to{\bf B}_n M$ decays,
both Cabibbo flavored (CF) and SCS decays are not well explained.
In particular, the experimental measurements show that
\begin{eqnarray}\label{data2}
{\cal B}_{p\pi^0}\equiv{\cal B} (\Lambda_c^+ \to p \pi^0)
&<&3\times10^{-4}\;(\text{90\% C.L.})\;\text{\cite{Ablikim:2017ors,ppi0}}\,,\nonumber\\
{\cal R}'_{K/\pi}\equiv
\frac{{\cal B}(\Xi^0_c \to \Xi^- K^+)}{{\cal B}(\Xi^0_c \to \Xi^- \pi^+)}
&=&0.028\pm 0.006\simeq(0.6\pm 0.2)s_c^2\;\text{\cite{pdg}}\,,\nonumber\\
{\cal B}(\Lambda_c^+ \to \Lambda^0 K^+)
&=&(6.1\pm 1.2)\times 10^{-4}\;\text{\cite{pdg}}\,,\nonumber\\
{\cal B}(\Lambda_c^+ \to \Sigma^0 K^+)
&=&(5.2\pm 0.8)\times 10^{-4}\;\text{\cite{pdg}}\,,
\end{eqnarray}
where $s_c\equiv \sin\theta_c=0.2248$~\cite{pdg}
with $\theta_c$   the well-known Cabbibo angle.
However, theoretical evaluations based on the $SU(3)_f$ symmetry lead to
${\cal B}_{p\pi^0}=(5.7\pm 1.5)\times 10^{-4}$ and
${\cal R}'_{K/\pi}\simeq 1.0 s_c^2$~\cite{Geng:2017esc},
and those in the factorization approach
give
${\cal B}_{p\pi^0}=f_{\pi}^2/(2f_K^2) s_c^2\,{\cal B}(\Lambda_c^+\to p\bar K^0)
=(5.5\pm 0.3)\times 10^{-4}$
and ${\cal R}'_{K/\pi}=(f_K/f_\pi)^2 s_c^2\simeq 1.4 s_c^2$,
where we have used the data of ${\cal B}(\Lambda_c^+\to p\bar K^0)=(3.16\pm 0.16)\times 10^{-2}$~\cite{pdg}.
In addition, the fitted results of
${\cal B}(\Lambda_c^+ \to \Lambda^0 K^+,\Sigma^0 K^+)
=(4.6\pm 0.9,4.0\pm 0.8)\times 10^{-4}$~\cite{Geng:2018plk}
are  $(1.3-1.6)\sigma$ away from the data in Eq.~(\ref{data2}).
%
In this study, we will consider  the breaking effects
of the $SU(3)_f$ symmetry due to the fact of $m_s\gg m_{u,d}$
in the ${\bf B}_c\to {\bf B}_n M$ decays, particularly,
the SCS processes,
in accordance with the $D\to MM$ ones.
%
Our goal is to find out whether the data in Eq.~(\ref{data2}) can be understood by introducing
the breaking effects.

The paper is organized as follows.
In  Sec. II, we provide the formalism, in which the amplitudes of the ${\bf B}_c\to {\bf B}_n M$ decays
with and without the breaking effects of $SU(3)_f$ symmetry are presented.
The numerical analysis is performed  in Sec. III.
In Sec. IV, we  discuss our results and give the conclusions.

\section{Formalism}

The two-body charmed baryon weak decays,
such as $\Xi^0_c \to \Xi^- \pi^+(\Xi^- K^+)$ and $\Lambda_c^+ \to p \pi^0$,
proceed through the quark-level
 transitions of $c\to  s u\bar d$, $c\to u d\bar d$ and $c\to u s\bar s$, with the
 effective Hamiltonian given by~\cite{Buras:1998raa}
\begin{eqnarray}\label{Heff}
{\cal H}_{eff}&=&\sum_{i=+,-}\frac{G_F}{\sqrt 2}c_i
\left(V_{cs}V_{ud}O_i+V_{cd}V_{ud} O_i^d+V_{cs}V_{us}O_i^s\right),
\end{eqnarray}
where $G_F$ is the Fermi constant,
 $c_{\pm}$ are the scale-dependent Wilson coefficients, and
the CKM matrix elements $V_{cs}V_{ud}\simeq 1$ and
$V_{cs}V_{us}\simeq-V_{cd}V_{ud}\simeq s_c$
correspond to
the Cabibbo-favored (CF) and singly Cabibbo-suppressed (SCS)
charmed hadron decays, respectively,
while $O^{(d,s)}_\pm$ are the four-quark operators,
 written as
\begin{eqnarray}\label{O12}
&&
O_\pm={1\over 2}\left[(\bar u d)(\bar s c)\pm (\bar s d)(\bar u c)\right]\,,\;\nonumber\\
&&
O_\pm^q={1\over 2}\left[(\bar u q)(\bar q c)\pm (\bar q q)(\bar u c)\right]\,,
\end{eqnarray}
where $q=(d,s)$ and $(\bar q_1 q_2)=\bar q_1\gamma_\mu(1-\gamma_5)q_2$.
With $q_i=(u,d,s)$ as the triplet of $3$,
the operator of $(\bar q^i q_k \bar q^j)c$ can be decomposed as
the irreducible forms, that is,
$(\bar 3\times 3\times \bar 3)c=(\bar 3+\bar 3'+6+\overline{15})c$.
Accordingly, the operators $O_-^{(q)}$ and $O_+^{(q)}$
belong to $6$ and $\overline{15}$, respectively, given by~\cite{Savage:1989qr}
\begin{eqnarray}
O_{\mp}\simeq &{\cal O}_{6(\overline{15})}&=\frac{1}{2}(\bar u d\bar s\mp\bar s d\bar u)c\,,\nonumber\\
O_{\mp}^q\simeq &{\cal O}_{6(\overline{15})}^q&=\frac{1}{2}(\bar u q\bar q\mp\bar q q\bar u)c\,,
\end{eqnarray}
such that  the effective Hamiltonian in Eq.~(\ref{Heff})
can be transformed into the tensor form of
\begin{eqnarray}\label{Heff2}
{\cal H}_{eff}&=&\frac{G_F}{\sqrt 2}\left[c_-  \frac{\epsilon^{ijl}}{2}H(6)_{lk}+c_+H(\overline{15})_k^{ij}\right]\,,
\end{eqnarray}
with the non-zero entries:
\begin{eqnarray}\label{nz_entry}
&&H_{22}(6)=2\,,H_2^{13}(\overline{15})=H_2^{31}(\overline{15})=1\,,\nonumber\\
&&H_{23}(6)=H_{32}(6)=-2s_c\,,
H^2_{12}(\overline{15})=H^2_{21}(\overline{15})=s_c\,,
\end{eqnarray}
where the notations of $(i,j,k)$ are quark indices,
to be connected to the initial and final states
in the amplitudes.
Note that $H_{23}(6)$ and $H_{32}(6)$ are derived from
${\cal O}^{s}_6$ and ${\cal O}^{d}_6$, respectively.
The lowest-lying charmed baryon states ${\bf B}_c$
are an anti-triplet of $\bar 3$
to consist of
$(ds-sd)c$, $(us-su)c$ and $(ud-du)c$,
presented as
\begin{eqnarray}
{\bf B}_{c}&=&(\Xi_c^0,-\Xi_c^+,\Lambda_c^+)\,,
\end{eqnarray}
together with the baryon and meson octets, given by
\begin{eqnarray}
{\bf B}_n&=&\left(\begin{array}{ccc}
\frac{1}{\sqrt{6}}\Lambda^0+\frac{1}{\sqrt{2}}\Sigma^0 & \Sigma^+ & p\\
 \Sigma^- &\frac{1}{\sqrt{6}}\Lambda^0 -\frac{1}{\sqrt{2}}\Sigma^0  & n\\
 \Xi^- & \Xi^0 &-\sqrt{\frac{2}{3}}\Lambda^0
\end{array}\right)\,,\nonumber\\
M&=&\left(\begin{array}{ccc}
\frac{1}{\sqrt{2}}\pi^0& \pi^- & K^-\\
 \pi^+ &-\frac{1}{\sqrt{2}}\pi^0& \bar K^0\\
 K^+ & K^0& 0
\end{array}\right)\,,
\end{eqnarray}
where we have removed the octet $\eta_8$ and singlet $\eta_1$ meson states to simplify our discussions.
%
Subsequently,
the amplitudes of ${\bf B}_c\to {\bf B}_n M$  can be derived as
\begin{eqnarray}\label{T-amp0}
{\cal A}({\bf B}_c\to {\bf B}_n M)
=\langle {\bf B}_n M|{\cal H}_{eff}|{\bf B}_c\rangle
=\frac{G_F}{\sqrt 2}T({\bf B}_{c}\to {\bf B}_nM)\,,
\end{eqnarray}
with $T({\bf B}_{c}\to {\bf B}_nM)=T({\cal H}_6)+T({\cal H}_{\overline{15}})$,
where
$T({\cal H}_{6,\overline{15}})$
are decomposed as~\cite{Geng:2017esc,Geng:2017mxn,Geng:2018plk}
\begin{eqnarray}\label{T-amp1}
T({\cal H}_6)&=&
a_1 H_{ij}(6)T^{ik}({\bf B}_n)_k^l (M)_l^j+
a_2 H_{ij}(6)T^{ik}(M)_k^l ({\bf B}_n)_l^j\nonumber\\
&+&
a_3 H_{ij}(6)({\bf B}_n)_k^i (M)_l^j T^{kl}
\,,
\nonumber\\
T({\cal H}_{\overline{15}})&=&
a_4H^{k}_{li}(\overline{15})({\bf B}_{c})^j (M)^i_j ({\bf B}_n)^l_k
+a_5({\bf B}_n)^i_j (M)^l_i H(\overline{15})^{jk}_l ({\bf B}_{c})_k\nonumber\\
&+&
a_6({\bf B}_n)^k_l (M)^i_j H(\overline{15})^{jl}_i ({\bf B}_{c})_k
+a_7({\bf B}_n)^l_i (M)^i_j H(\overline{15})^{jk}_l ({\bf B}_{c})_k\,,
\end{eqnarray}
with $T^{ij} \equiv ({\bf B}_c)_k\epsilon^{ijk}$.
%
In Eq.~(\ref{T-amp1}), $a_{1,2,3}$ and $a_{4,5,6,7}$
are the $SU(3)$ parameters from $H(6)$ and $H(\overline{15})$, respectively,
in which $c_{\mp}$ have been absorbed.
We hence obtain~\cite{Geng:2018plk}
\begin{eqnarray}\label{T-amp1a}
T(\Lambda_c^+\to p \pi^{0})&=&-\sqrt 2(a_{2}+a_3-\frac{a_{6} - a_{7}}{2})s_c\,,\nonumber\\
T(\Xi_c^0\to\Xi^{-} \pi^{+})&=&2(a_{1}+\frac{a_{5} + a_{6}}{2})\,,\nonumber\\
T(\Xi_c^0\to\Xi^{-} K^{+})&=&-2(a_{1}+\frac{a_{5} + a_{6}}{2})s_c\,,
\end{eqnarray}
based on the exact $SU(3)_f$ symmetry.

According to Refs.~\cite{Geng:2017esc,Geng:2018plk},
the numerical analysis with the minimum $\chi^2$ fit
has well explained the ten observed ${\bf B}_c\to {\bf B}_nM$ decays
by neglecting the terms associated with $a_{4,5,6,7}$
to reduce the parameters~\cite{Geng:2017esc,Geng:2018plk,
Geng:2017mxn,Lu:2016ogy,Wang:2017gxe}.
This reduction is due to the fact that
the contributions  to the branching rates from $H(\overline{15})$ and $H(6)$
lead to a small ratio of  ${\cal R}(\overline{15}/6)=c_+^2/c_-^2\simeq 17\%$
with $(c_+,c_-)=(0.76,1.78)$ calculated at the scale $\mu=1$ GeV
in the NDR scheme~\cite{Fajfer:2002gp,Li:2012cfa}.
There remain two measurements to be explained.
In Eq.~(\ref{data2}), the prediction  for ${\cal B}_{p\pi^0}$ has the 2$\sigma$ gap
to reach the edge of the experimental upper bound.
However, with ${\cal R}(\overline{15}/6)$ 
to be small, 
it is nearly impossible that,
by restoring $a_{4,5,6,7}$ that have been ignored
in the literature~\cite{Geng:2017esc,Geng:2018plk,
Geng:2017mxn,Lu:2016ogy,Wang:2017gxe},
one can accommodate the data of ${\cal B}_{p\pi^0}$ 
but without having impacts on the other decay modes,
which are correlated with the same sets of parameters.
Moreover, as seen from Eq.~(\ref{T-amp1a}),
there is no room for ${\cal R}'_{K/\pi}$ as it is
 fixed to be $ (1.0)s_c^2$.
On the other hand, the results for
$D\to MM$ decays in Eq.~(\ref{data1}) suggest some possible corrections from
the breaking effects of the $SU(3)_f$ symmetry
in the SCS processes.
%
In the charm baryon decays, we consider the similar effects.
Due to $m_s\gg m_{u,d}$, we present
the matrix of $M_s=\epsilon (\lambda_s)^i_j$~\cite{Grossman:2012ry}
to break  $SU(3)_f$, where $\epsilon\sim 0.2-0.3$ and
$\lambda_s$ is given by~\cite{Grossman:2012ry,Pirtskhalava:2011va,Savage:1991wu}
\begin{eqnarray}
\lambda_s
&=&\left(\begin{array}{ccc}
1&0&0\\
0&1&0\\
0&0&-2
\end{array}\right)\,,
\end{eqnarray}
which transforms as an octet of 8, such that its coupling to $H(6)$
is in the form of $8\times 6=\bar 3+6+\overline{15}+24$,
and $\bar 3$ is for 
the simplest break effects 
to be confined in the SCS processes~\cite{Savage:1991wu,Abbott:1979fw}.
%
Note that from
$\frac{1}{8}\left(\delta^i_l(\lambda_s)^n_jH(6)_{kn}-\delta^i_l(\lambda_s)^n_kH(6)_{jn}+\delta^i_k(\lambda_s)^n_jH(6)_{ln}-\delta^i_k(\lambda_s)^n_lH(6)_{jn}\right)$  and
the nonzero entry of $H(\bar 3)^1=s_c$
from the coupling of $H(6)_{23}$ and $H(6)_{32}$~\cite{Grossman:2012ry}, one can trace back to
the break effect between SCS $c\to us\bar s$ and $c\to ud\bar d$ transitions.
As a result, the $SU(3)_f$ symmetry breaking gives rise to the new $T$-amplitudes,
given by
\begin{eqnarray}\label{T-amp2}
T({\cal H}_3)
&=&
v_1({\bf B}_c)_iH(\overline{3})^i({\bf B}_n)^j_k(M)^k_j+
v_2({\bf B}_c)_i H(\overline{3})^j({\bf B}_n)^i_k(M)^k_j\nonumber\\
&+&
v_3({\bf B}_c)_i H(\overline{3})^j({\bf B}_n)^k_j(M)^i_k\,,
\end{eqnarray}
where  $v_{1,2,3}$ are the parameters related to
the  $SU(3)_f$ breaking.
%
It is interesting to note that the $v_1$ terms  associated with
$({\bf B}_c)_iH(\overline{3})^i$ in Eq.~(\ref{T-amp2})   occur in some of the $\Xi_c^{0,+}$ decays, but disappear in all
 $\Lambda_c^+$ modes.
%
By adding $T({\cal H}_3)$ into $T({\bf B}_c\to{\bf B}_n M)$ in Eq.~(\ref{T-amp0}),
the full expansions of $T({\bf B}_c\to{\bf B}_n M)$ are given in Table~\ref{tab_Tamp},
to be used to calculate the decay widths, given by~\cite{pdg}
\begin{eqnarray}
\Gamma({\bf B}_c\to {\bf B}_n M)=
\frac{|\vec{p}_{cm}|}{8\pi m_{{\bf B}_c}^2}|{\cal A}({\bf B}_c\to {\bf B}_n M)|^2\,,
\end{eqnarray}
where $|\vec{p}_{cm}|=
\sqrt{[(m_{{\bf B}_c}^2-(m_{{\bf B}_n}+m_M)^2]
[(m_{{\bf B}_c}^2-(m_{{\bf B}_n}-m_M)^2]}/(2 m_{{\bf B}_c})$.

%
\begin{table}[t!]
\caption{Amplitudes of $T({\bf B}_{c}\to {\bf B}_n M)$, where
 $T$-amps referes to $T({\bf B}_{c}\to {\bf B}_n M)$ and
 CF (SCS) represents Cabbibo favored (singly Cabbibo-suppressed).
 }\label{tab_Tamp}
{
\scriptsize
\begin{tabular}{|l|l|}
\hline
CF mode&\;\;\;\;\;\;\;\;$T$-amp
\\
\hline
$\Xi_c^0\to\Sigma^{+} K^{-} $
& $ 2(a_{2}+\frac{a_{4} + a_{7}}{2})$
\\
$\Xi_c^0\to\Sigma^{0}\bar{K}^{0}$
&$-\sqrt{2}(a_{2}+a_{3}$
$-\frac{a_{6}-a_{7}}{2})$
\\
$\Xi_c^0\to\Xi^{0} \pi^{0} $
& $ -\sqrt{2}(a_{1}-a_{3}$
$-\frac{a_{4}-a_{5}}{2})$
\\
$\Xi_c^0\to\Xi^{-} \pi^{+} $
& $ 2(a_{1}+\frac{a_{5} + a_{6}}{2})$
\\
$\Xi_c^0\to\Lambda^{0} \bar{K}^{0} $
&
$-\sqrt{\frac{2}{3}}(2a_1-a_2-a_3$
$+\frac{2a_5-a_6-a_7}{2})$\\[15mm]
\hline
$\Xi_c^+\to\Sigma^{+} \bar{K}^{0} $
&$ -2(a_{3}-\frac{a_{4} + a_{6}}{2})$
\\
$\Xi_c^+\to\Xi^{0} \pi^{+} $
& $2(a_{3}+\frac{a_{4} + a_{6}}{2})$
\\[15mm]
\hline
$\Lambda_c^+\to\Sigma^{0} \pi^{+} $
&$-\sqrt{2}(a_1-a_2-a_3$
$-\frac{a_5-a_7}{2})$
\\
$\Lambda_c^+\to\Sigma^{+} \pi^{0} $
& $\sqrt{2}(a_{1}-a_{2}-a_{3}$
$-\frac{a_{5}-a_{7}}{2})$
\\
$\Lambda_c^+\to\Xi^{0} K^{+} $
& $-2(a_{2}-\frac{a_{4} + a_{7}}{2})$
\\
$\Lambda_c^+\to p \bar{K}^{0} $
& $ -2(a_{1}-\frac{a_{5} + a_{6}}{2})$
\\
$\Lambda_c^+\to\Lambda^{0} \pi^{+} $
&
$-\sqrt{\frac{2}{3}}(a_1+a_2+a_3-\frac{a_5-2a_6+a_7}{2})$
\\
\hline
\end{tabular}
\begin{tabular}{|l|l|}
\hline
SCS mode&\;\;\;\;\;\;\;\;$T$-amp
\\\hline
$\Xi_c^0\to\Sigma^{+} \pi^{-} $
& $2(a_{2}+v_1+v_3+\frac{a_{4} + a_{7}}{2})s_c$
\\
$\Xi_c^0\to\Sigma^{-} \pi^{+} $
& $2(a_{1}+v_1+v_2+\frac{a_{5} + a_{6}}{2})s_c$
\\
$\Xi_c^0\to\Sigma^{0} \pi^{0} $
&$(a_{2}+a_{3}-2v_1-v_2-v_3$
$-\frac{a_{4}-a_{5}+a_{6}-a_{7}}{2})s_c$
\\
$\Xi_c^0\to\Xi^{-} K^{+} $
& $ -2(a_{1}-v_1-v_2+\frac{a_{5} + a_{6}}{2})s_c$
\\
$\Xi_c^0\to p K^-$
&$-2(a_{2}-v_1-v_3+\frac{a_{4} + a_{7}}{2})s_c$
\\
$\Xi_c^0\to\Xi^{0} K^{0} $
& $-2(a_{1}-a_2-a_{3}-v_1+\frac{a_{5}-a_{7}}{2})s_c$
\\
$\Xi_c^0\to n \bar K^{0} $
& $2(a_{1}-a_{2}-a_{3}+v_1+\frac{a_{5}-a_{7}}{2})s_c$
\\
$\Xi_c^0\to\Lambda^{0} \pi^{0} $
&
$\sqrt{\frac{1}{3}}(-a_1-a_2+2a_3+v_2+v_3+\frac{a_4-a_5-a_6-a_7}{2})s_c$
\\
\hline
$\Xi_c^+\to\Sigma^{0} \pi^{+} $
&$-\sqrt 2(a_{1}-a_{2}+v_2-v_3+\frac{a_{4}-a_{5}+a_{6}+a_{7}}{2})s_c$
\\
$\Xi_c^+\to\Sigma^{+} \pi^{0} $
&$\sqrt 2(a_{1}-a_{2}+v_2-v_3-\frac{a_{4}+a_{5}+a_{6}-a_{7}}{2})s_c$
\\
$\Xi_c^+\to\Xi^{0} K^{+} $
& $2(a_2+a_{3}-v_2+\frac{a_{6} - a_{7}}{2})s_c$
\\
$\Xi_c^+\to p \bar K^{0} $
& $2(a_1-a_{3}-v_3+\frac{a_{4} - a_{5}}{2})s_c$
\\
$\Xi_c^+\to\Lambda^0\pi^+$
& $\sqrt{\frac{2}{3}}(a_1+a_2-2a_3-v_2-v_3-\frac{3a_4+a_5+a_6+a_7}{2})s_c$
\\
\hline
$\Lambda_c^+\to\Sigma^{+} K^{0} $
& $2(a_{1}-a_{3}+v_3-\frac{a_{4}-a_{5}}{2})s_c$
\\
$\Lambda_c^+\to\Sigma^{0} K^{+} $
&$\sqrt{2}(a_1-a_3+v_3-\frac{a_4+a_5}{2})s_c$
\\
$\Lambda_c^+\to p \pi^{0} $
& $ \sqrt 2(a_{2}+a_3+v_2-\frac{a_{6} - a_{7}}{2})s_c$
\\
$\Lambda_c^+\to n\pi^+$
&$2(a_{2}+a_3+v_2+\frac{a_{6} - a_{7}}{2})s_c$
\\
$\Lambda_c^+\to\Lambda^{0} K^{+} $
&
$\sqrt{\frac{2}{3}}(a_1-2a_2+a_3-2v_2+v_3-\frac{3a_4-a_5+2a_6+2a_7}{2})s_c$\\
\hline
\end{tabular}
}
\end{table}

\section{Numerical Analysis} 

In the numerical analysis, we examine
${\cal B}(\Lambda_c^+ \to \Lambda^0 K^+,\Sigma^0 K^+,p\pi^0)$ and ${\cal R}'_{K/\pi}$
by including the breaking effects of the $SU(3)_f$ symmetry
to see if one can explain their data  in Eq.~(\ref{data2}).
%
%
The theoretical inputs for the CKM matrix elements are given by~\cite{pdg}
\begin{eqnarray}\label{B1}
&&(V_{cs},V_{ud},V_{us},V_{cd})=(1-\lambda^2/2,1-\lambda^2/2,\lambda,-\lambda)\,,
\end{eqnarray}
with $\lambda=0.2248$ in the Wolfenstein parameterization.
We perform the minimum $\chi^2$ fit,
in terms of the equation of~\cite{Geng:2018plk}
\begin{eqnarray}
\chi^2=
\sum_{i} \bigg(\frac{{\cal B}^i_{th}-{\cal B}^i_{ex}}{\sigma_{ex}^i}\bigg)^2+
\sum_{j}\bigg(\frac{{\cal R}^j_{th}-{\cal R}^j_{ex}}{\sigma_{ex}^j}\bigg)^2\,,
\end{eqnarray}
with ${\cal B}={\cal B}(\Lambda_c^+\to {\bf B}_nM)$ and
${\cal R}={\cal B}(\Xi_c^0\to {\bf B}_nM)/{\cal B}(\Xi_c^0\to \Xi^-\pi^+)$,
where the subscripts $th$ and $ex$ are denoted as
the theoretical inputs from the amplitudes in Table~\ref{tab_Tamp} and
the experimental data points in Table~\ref{data}, respectively,
while $\sigma_{i,j}$ correspond to the 1$\sigma$ errors.
\begin{table}[t!]
\caption{The data of the ${\bf B}_c\to {\bf B}_n M$ decays,
together with the reproduction with the exact (broken) $SU(3)_f$ symmetry
in the 3rd (4th) column.}\label{data}
\begin{tabular}{|c||c|c|c|}
\hline
(Branching) Ratios
&Data~\cite{pdg,
Ablikim:2017ors,Ablikim:2018bir}
&Exact ~\cite{Geng:2018plk}&Broken \\
\hline
$10^4{\cal B}(\Lambda_c^+ \to \Sigma^0 K^+)$
&$5.2\pm 0.8$
&$4.0\pm 0.8$
&$5.2\pm 0.7$\\
$10^4{\cal B}(\Lambda_c^+ \to \Lambda^0 K^+)$
&$6.1\pm 1.2$
&$4.6\pm 0.9$
&$6.1\pm 0.9$\\
%
%
%
${\cal R}'_{K/\pi}=
\frac{{\cal B}(\Xi^0_c \to \Xi^- K^+)}{{\cal B}(\Xi^0_c \to \Xi^- \pi^+)}$
&$(0.6\pm 0.2)s_c^2$
&$(1.0)s_c^2$
&$(0.6\pm 0.2)s_c^2$\\
\hline
$10^2{\cal B}(\Lambda_c^+ \to \Sigma^0 \pi^+)$
& $1.29\pm 0.07$
& $1.3\pm 0.2$
& $1.3\pm 0.1$\\
$10^2{\cal B}(\Lambda_c^+ \to \Sigma^+ \pi^0)$
&$1.24\pm 0.10$
&$1.3\pm 0.2$
&$1.3\pm 0.1$\\
$10^2{\cal B}(\Lambda_c^+ \to \Xi^0 K^+)$
& $0.59\pm 0.09$
& $0.5\pm 0.1$
& $0.6\pm 0.1$\\
$10^2{\cal B}(\Lambda_c^+ \to p \bar K^0)$
&$3.16\pm 0.16$
&$3.3\pm 0.2$
&$3.2\pm 0.1$\\
$10^2{\cal B}(\Lambda_c^+ \to \Lambda^0 \pi^+)$
&$1.30\pm 0.07$
&$1.3\pm 0.2$
&$1.3\pm 0.1$\\
%
${\cal R}''_{K/\pi}=\frac{{\cal B}( \Xi^0_c \to \Lambda^0 \bar K^0)}{{\cal B}(\Xi^0_c \to \Xi^-\pi^+)}$
&$0.42\pm 0.06$
&$0.5\pm 0.1$
&$0.5\pm 0.1$\\
\hline
\end{tabular}
\end{table}
%
By following~Refs.~\cite{Geng:2017esc,Geng:2018plk,Geng:2017mxn},
we extract the parameters, which are in fact complex numbers, given by
\begin{eqnarray}\label{8p}
&&a_1, a_2e^{i\delta_{a_2}},a_3e^{i\delta_{a_3}},
v_1e^{i\delta_{v_1}}, v_2e^{i\delta_{v_2}},v_3e^{i\delta_{v_3}}\,,
\end{eqnarray}
where $a_{4,5, ...,7}$ have been ignored as discussed in Sec.~III.
Since only the relative phases contribute to the branching ratios,
$a_1$ is set to be real without losing generality.
However, we take  $v_i$
to be real numbers in order to fit 8 parameters
with the 9 data points in Table~\ref{data}.
%
In the calculation,
 $\delta_{a_i}$ (i=2 or 3) from 
 $a_i e^{\delta_{a_i}}$ is a fitting parameter, which can absorb
 the phase of $\delta_{v_i}$ from the interference in the data fitting.
Note that $\delta_{a_i}\ (i=2,3)$ have been fitted with the imaginary parts~\cite{Geng:2018plk}.
As a result, we may set $\delta_{v_{2,3}}$ along with the overall phase of $\delta_{v_1}$ to be zero
for the estimations of the decay branching ratios due to the $SU(3)$ breaking effects.
We will follow Ref.~\cite{Wang:2017gxe} to test our assumption, where a similar global fit
in the approach of the $SU(3)_f$ symmetry has been done to
extract $a_2 e^{\delta_{a_2}}$ by
 freely rotating the angle of $\delta_{a_2}$ from $-180^\circ$ to $180^\circ$
to estimate the uncertainties of the branching ratios.
Subsequently, the fit with the breaking effects in the  $SU(3)_f$ symmetry
yields
\begin{eqnarray}\label{su3_fit}
&&(a_1,a_2,a_3)=(0.252\pm 0.005,0.127\pm 0.009,0.091\pm 0.015)\,\text{GeV}^3\,,\nonumber\\
&&(\delta_{a_2},\delta_{a_3})=(73.0\pm 27.3, 40.2\pm 4.7)^\circ\,,\nonumber\\
&&(v_1,v_2,v_3)=(0.090\pm 0.032,-0.037\pm 0.013,0.025\pm 0.012)\,\text{GeV}^3\,,
\end{eqnarray}
with $\chi^2/d.o.f=3.0/1$, where $d.o.f$ represents the degree of freedom.
Note that $a_{1,2,3}$ and their phases are nearly the same as those
without the breaking of
$SU(3)_f$~\cite{Geng:2018plk}.
With the parameters in Eq.~(\ref{su3_fit}),
we obtain the branching ratios of the CF and SCS ${\bf B}_c\to {\bf B}_nM$ decays,
shown in Table~\ref{tab_br}.

\begin{table}[t!]
\caption{The branching ratios of the ${\bf B}_{c}\to {\bf B}_n M$ decays,
where the numbers  with the dagger ($\dagger$)
correspond to the reproductions of the experimental data input,
instead of the predictions.}\label{tab_br}
{
\begin{tabular}{|c||c|c|}
\hline
CF mode&Exact~\cite{Geng:2018plk}&Broken\\
\hline
$ 10^3{\cal B}(\Xi_{c}^{0}  \to  \Sigma^{+} K^{-}) $ & $ 3.5 \pm 0.9 $ &$ 3.8 \pm 0.6 $ \\
$ 10^3{\cal B}(\Xi_{c}^{0}  \to  \Sigma^{0} \bar{K}^{0}) $ & $ 4.7 \pm 1.2 $ &$ 5.2 \pm 0.8 $ \\
$ 10^3{\cal B}(\Xi_{c}^{0}  \to  \Xi^{0} \pi^{0}) $ & $ 4.3 \pm 0.09 $ &$ 4.4 \pm 0.4 $ \\
$ 10^3{\cal B}(\Xi_{c}^{0}  \to  \Xi^{-} \pi^{+}) $ & $ 15.7 \pm 0.7 $ &$ 15.2 \pm 0.7 $ \\
$ 10^{\cal B}(\Xi_{c}^{0}  \to  \Lambda^{0} \bar{K}^{0} )$ & $ 8.3 \pm 0.9 $ &$ 7.8 \pm 0.5 $ \\
[23mm]
\hline
$ 10^3{\cal B}(\Xi_{c}^{+}  \to  \Sigma^{+} \bar{K}^{0} )$ & $ 8.0 \pm 3.9 $ &$ 7.8 \pm 2.7 $ \\
$ 10^3{\cal B}(\Xi_{c}^{+}  \to  \Xi^{0} \pi^{+}) $ & $ 8.1 \pm 4.0 $ &$ 7.9 \pm 2.7 $ \\
[23mm]
\hline
$10^2{\cal B}( \Lambda_{c}^{+}  \to  \Sigma^{0} \pi^{+} )$ & $(1.3 \pm 0.2)^\dagger $ &$ (1.3 \pm 0.1)^\dagger $ \\
$10^2{\cal B}( \Lambda_{c}^{+}  \to  \Sigma^{+} \pi^{0} )$ & $ (1.3 \pm 0.2)^\dagger $ &$(1.3 \pm 0.1)^\dagger $ \\
$10^2{\cal B}( \Lambda_{c}^{+}  \to  \Xi^{0} K^{+} )$ &$ (0.5\pm 0.1)^\dagger $ &$ (0.6\pm 0.1)^\dagger $ \\
$10^2{\cal B}( \Lambda_{c}^{+}  \to  p \bar{K}^{0} $ )& $ (3.3 \pm 0.2)^\dagger $ &$ (3.2 \pm 0.1)^\dagger $ \\
$10^2{\cal B}( \Lambda_{c}^{+}  \to  \Lambda^{0} \pi^{+} )$& $ (1.3 \pm 0.2)^\dagger $ &$ (1.3 \pm 0.1)^\dagger $ \\
\hline
\end{tabular}
\begin{tabular}{|c||c|c|}
\hline
SCS mode&Exact~\cite{Geng:2018plk}
&Broken\\
\hline
$10^4{\cal B}( \Xi_{c}^{0}  \to  \Sigma^{+} \pi^{-} )$ &$ 2.0 \pm 0.5 $&$ 5.2 \pm 1.6 $ \\
$10^4{\cal B}( \Xi_{c}^{0}  \to  \Sigma^{-} \pi^{+} )$ &$ 9.0 \pm 0.4 $&$ 12.8 \pm 3.1 $ \\
$10^4{\cal B}( \Xi_{c}^{0}  \to  \Sigma^{0} \pi^{0} )$ &$ 3.2 \pm 0.3 $&$ 7.7 \pm 2.2 $ \\
$10^4{\cal B}( \Xi_{c}^{0}  \to  \Xi^{-} K^{+} )$ &$ 7.6 \pm 0.4 $&$ 4.6 \pm 1.7 $ \\
$10^4{\cal B}( \Xi_{c}^{0}  \to  p K^{-} )$ &$ 2.1 \pm 0.5 $&$ 3.0 \pm 1.0 $ \\
$10^4{\cal B}( \Xi_{c}^{0}  \to  \Xi^{0} K^{0} )$ &$ 6.3 \pm 1.2 $&$ 4.1 \pm 0.7 $ \\
$10^4{\cal B}( \Xi_{c}^{0}  \to  n \bar{K}^{0} )$ &$ 7.9 \pm 1.4 $&$ 12.7 \pm 2.4 $ \\
$10^4{\cal B}( \Xi_{c}^{0}  \to  \Lambda^{0} \pi^{0} )$ &$ 0.2 \pm 0.2 $&$ 0.3 \pm 0.1 $ \\
\hline
$10^4{\cal B}( \Xi_{c}^{+}  \to  \Sigma^{0} \pi^{+} )$ &$ 18.5 \pm 2.2 $&$ 10.3 \pm 1.7 $ \\
$10^4{\cal B}( \Xi_{c}^{+}  \to  \Sigma^{+} \pi^{0} )$ &$ 18.5 \pm 2.2 $&$ 10.3 \pm 1.7 $ \\
$10^4{\cal B}( \Xi_{c}^{+}  \to  \Xi^{0} K^{+} )$ &$ 18.0 \pm 4.7 $&$ 24.3 \pm 4.1 $ \\
$10^4{\cal B}( \Xi_{c}^{+}  \to  p \bar{K}^{0} )$ &$ 20.3 \pm 4.2 $&$ 16.1 \pm 2.8 $ \\
$10^4{\cal B}( \Xi_{c}^{+}  \to  \Lambda^{0} \pi^{+} )$ &$ 1.6 \pm 1.2 $&$ 2.4 \pm 1.0 $ \\
\hline
$10^4{\cal B}( \Lambda_{c}^{+}  \to  \Sigma^{+} K^{0} )$ &$ 8.0 \pm 1.6 $&$ 10.4 \pm 1.5 $ \\
$10^4{\cal B}( \Lambda_{c}^{+}  \to  \Sigma^{0} K^{+} )$ &$ (4.0\pm 0.8)^\dagger$&$ (5.2\pm 0.7)^\dagger $ \\
$10^4{\cal B}( \Lambda_{c}^{+}  \to  p \pi^{0} )$ &$ 5.7 \pm 1.5 $&$ 5.4 \pm 1.0 $ \\
$10^4{\cal B}( \Lambda_{c}^{+}  \to  n \pi^{+} )$ &$ 11.3 \pm 2.9 $&$ 10.7 \pm 1.9 $ \\
$10^4{\cal B}( \Lambda_{c}^{+}  \to  \Lambda^{0} K^{+} )$ &$ (4.6 \pm 0.9)^\dagger $&$ (6.1 \pm 0.9)^\dagger $ \\
\hline
\end{tabular}
}
\end{table}
%

\section{Discussions and Conclusions}
As seen from Tables~\ref{data} and \ref{tab_br},
the breaking effects associated with $v_2$ and $v_3$ on the branching ratios of the SCS $\Lambda_c^+\to {\bf B}_nM$ decays
are  at most around $30\%$,
%
which is close to the naive estimation of
$(f_K/f_\pi)^2\simeq 40\%$.
In particular, we get
${\cal B}(\Lambda_c^+ \to \Lambda^0 K^+,\Sigma^0 K^+)
=(6.1\pm 0.9,5.2\pm 0.7)\times 10^{-4}$, which explain the data in Eq.~(\ref{data2}) well and
alleviate the $(1.3-1.6)\sigma$ deviations by the fit with the exact $SU(3)_f$ symmetry~\cite{Geng:2018plk}.
Meanwhile,
the  branching ratios for the CF modes are fitted to be the same as those without the breaking
except  ${\cal B}(\Lambda_c^+ \to \Xi^0 K^+)$,
which is slightly different in order to account for the recent observational value given by BESIII~\cite{Ablikim:2018bir}.
%
%

Moreover, the fitted value of ${\cal R}'_{K/\pi}=(0.6\pm 0.2)s_c^2=0.03\pm 0.01$
explains the data very well for the first time. This leads to the prediction of
${\cal B}(\Xi_{c}^{0}  \to  \Xi^{-} K^{+})=(4.6 \pm 1.7)\times 10^{-4}$,
with $v_1+v_2$ as the destructive contribution
to reduce the value of  $(7.6 \pm 0.4)\times 10^{-4}$ under the exact $SU(3)_f$ symmetry,
whereas ${\cal B}( \Xi_{c}^{0}\to\Sigma^{-} \pi^{+})
=(12.8 \pm 3.1)\times 10^{-4}$ receives the constructive contribution
from $v_1+v_2$, with $T(\Xi_c^0\to \Xi^{-} K^{+},\Sigma^-\pi^+)=\mp [a_1\mp (v_1+v_2)]s_c$.
Since there are other similar interferences between $a_i$ and $v_i$,
which come from $T(\Xi_c^0\to pK^-,\Sigma^+\pi^-)=\mp [a_2\mp (v_1+v_3)]s_c$ and
$T(\Xi_c^+\to \Sigma^0 \pi^+,\Sigma^+\pi^0)=\mp \sqrt 2 [(a_1-a_2)+(v_2-v_3)]s_c$,
it is predicted that
${\cal B}(\Xi_c^0\to pK^-,\Sigma^+\pi^-)=(3.0 \pm 1.0, 5.2 \pm 1.6)\times 10^{-4}$ and
${\cal B}(\Xi_c^+\to \Sigma^0 \pi^+(\Sigma^+\pi^0))=( 10.3 \pm 1.7)\times 10^{-4}$.
It is  interesting to note that
the  important roles of the terms associated with
$v_1$ in the $T$ amplitudes are also projected in the $\Xi_c^{0,+}$ modes, particularly,  $\Xi_c^0\to\Xi^{-} K^{+}$
and $ \Xi_{c}^{0}\to\Sigma^{0} \pi^{0}$. Clearly, these SCS $\Xi_c$ decays all contain sizable $SU(3)_f$ breaking effects, and
can be treated as golden modes to test  the $SU(3)_f$ symmetry.

In our calculation, we treat $v_3$ as the norm
in $T(\Lambda_c^+\to\Sigma^{0} K^{+})\simeq \sqrt{2}(a_1-a_3+v_3)s_c$ of Table~I,
such that $\delta_{v_3}$ is allowed to rotate from $-90^\circ$ to $50^\circ$
without letting ${\cal B}( \Lambda_{c}^{+}  \to  \Sigma^{0} K^{+})$ exceed the data.
Since the allowed range for $\delta_{v_{3}}$ is large, it is clear that
its value is insensitive  to the data.
On the other hand, in order to explain the experimental data of ${\cal R}'_{K/\pi}$ 
  with the smallest corrections from  $v_i e^{\delta_{v_i}}$,
   we assume maximumly destructive interferences between $a_i e^{\delta_{a_i}}$ and $v_i e^{\delta_{v_i}}$. 
   Explicitly, in $T(\Xi_c^0\to\Xi^{-} K^{+})\simeq -2(a_1-v_1-v_2)s_c$ for  ${\cal R}'_{K/\pi}$, 
    we can take $\delta_{v_1}=\delta_{v_2}=\delta_{a_1}=0$ as an overall phase in $T(\Xi_c^0\to\Xi^{-} K^{+})$.
Consequently, we are able to assume real values for  $v_i$ (i=1,2,3)  without loss of generality.
Finally, we remark that, even with the breaking effects, we are still unable to fit the data of
$\Lambda_c^+\to p\pi^0$ in Eq.~(\ref{data1}) as our result for its branching ratio  of
$(5.4\pm 1.0)\times 10^{-4}$, which is close to $(5.5\pm 0.3)\times 10^{-4}$ from the factorization approach~\cite{Geng:2018plk},
is lower than the current
experimental upper bound of $3\times 10^{-4}$~\cite{Ablikim:2017ors,ppi0}.
However, it is possible that 
$H(\overline{15})$ would be  non-negligible in $\Lambda_c^+\to p\pi^0$. 
For example, with $T(\Lambda_c^+\to p\pi^0)
=\sqrt 2(a_{2}+a_3+v_2-(a_{6} - a_{7})/2)s_c$ in Table~\ref{tab_Tamp},
the contribution from $(a_6-a_7)/2$ of $H(\overline{15})$ 
might be comparable with that from $a_2+a_3+v_2$ of $H(6)$,
while $a_{2,3}$ and $v_2$ of Eq.~(\ref{su3_fit}) are taken to be small.
In particular, with $(a_6-a_7)/2$ to be around $25\%$ of $a_{2}+a_3+v_2$,
${\cal B}(\Lambda_c^+\to p\pi^0)$ can be reduced to be 
within the experimental upper bound due to the destructive interference. 
In this case, there is a corresponding constructive interference in $\Lambda_c^+\to n\pi^+$,
leading to ${\cal B}(\Lambda_c^+\to n\pi^+)\sim 17\times 10^{-4}$, which breaks the relation of 
${\cal A}(\Lambda_c^+\to n\pi^+)=
\sqrt 2 {\cal A}(\Lambda_c^+\to p\pi^0)$~\cite{Cheng:2018hwl}.
Clearly, in order to confirm the importance of $H(\overline{15})$, both experimental observations  of 
$\Lambda_c^+\to p\pi^0$ and $\Lambda_c^+\to n\pi^+$ are needed.

In sum, we have studied the singly Cabibbo-suppressed charmed baryon decays.
We have shown that the breaking effect of the $SU(3)_f$ symmetry can be used to understand
the experimental data of ${\cal B}(\Lambda_c^+\to \Sigma^{0} K^{+},\Lambda^{0} K^{+})$ and
$R'_{K/\pi}$=${\cal B}(\Xi^0_c \to \Xi^- K^+)$/${\cal B}(\Xi^0_c \to \Xi^- \pi^+)$.
With these effects, we
 have  obtained  that
${\cal B}(\Xi_{c}^{0}  \to  \Xi^{-} K^{+},\Sigma^{-} \pi^{+})
=(4.6 \pm 1.7,12.8 \pm 3.1)\times 10^{-4}$,
${\cal B}(\Xi_c^0\to pK^-,\Sigma^+\pi^-)
=(3.0 \pm 1.0, 5.2 \pm 1.6)\times 10^{-4}$ and
${\cal B}(\Xi_c^+\to \Sigma^0 \pi^+(\Sigma^+\pi^0))=(10.3 \pm 1.7)\times 10^{-4}$,
which are quite different from those predicted by the approach with the exact $SU(3)_f$ symmetry.
However, even with the breaking effects, our result  for the branching ratio of $\Lambda_c^+\to p\pi^0$
is still higher than the current  experimental upper bound,
which clearly requires a close examination by a future dedicated experiment.

\section*{ACKNOWLEDGMENTS}
This work was supported in part by National Center for Theoretical Sciences,
MoST (MoST-104-2112-M-007-003-MY3), and
National Science Foundation of China (11675030).


\begin{thebibliography}{99}
\bibitem{Khodjamirian:2017zdu}
A.~Khodjamirian and A.A.~Petrov,
Phys.\ Lett.\ B {\bf 774}, 235 (2017). 

\bibitem{Cheng:1991sn}
H.Y.~Cheng and B.~Tseng,
Phys.\ Rev.\ D {\bf 46}, 1042 (1992); {\bf 55}, 1697(E) (1997).

\bibitem{Cheng:1993gf}
H.Y.~Cheng and B.~Tseng,
Phys.\ Rev.\ D {\bf 48}, 4188 (1993).

\bibitem{Zenczykowski:1993hw}
P.~Zenczykowski,
Phys.\ Rev.\ D {\bf 50}, 402 (1994).


\bibitem{Fayyazuddin:1996iy}
Fayyazuddin and Riazuddin,
Phys.\ Rev.\ D {\bf 55}, 255; {\bf 56}, 531(E) (1997).


\bibitem{Dhir:2015tja}
R.~Dhir and C.S.~Kim,
Phys.\ Rev.\ D {\bf 91}, 114008 (2015). 

\bibitem{Cheng:2018hwl}
H.Y.~Cheng, X.W.~Kang and F.~Xu,
Phys.\ Rev.\ D {\bf 97}, 074028 (2018).
 


\bibitem{He:2000ys}
X.G.~He, Y.K.~Hsiao, J.Q.~Shi, Y.L.~Wu and Y.F.~Zhou,
Phys.\ Rev.\ D {\bf 64}, 034002 (2001).

\bibitem{Fu:2003fy}
H.K.~Fu, X.G.~He and Y.K.~Hsiao,
Phys.\ Rev.\ D {\bf 69}, 074002 (2004).

\bibitem{Hsiao:2015iiu}
Y.K.~Hsiao, C.F.~Chang and X.G.~He,
Phys.\ Rev.\ D {\bf 93}, 114002 (2016).

\bibitem{He:2015fwa}
X.G.~He and G.N.~Li,
Phys.\ Lett.\ B {\bf 750}, 82 (2015).

\bibitem{He:2015fsa}
M.~He, X.G.~He and G.N.~Li,
Phys.\ Rev.\ D {\bf 92}, 036010 (2015).

\bibitem{Grossman:2012ry}
Y.~Grossman and D.J.~Robinson, 
JHEP {\bf 1304}, 067 (2013). 

\bibitem{Pirtskhalava:2011va}
D.~Pirtskhalava and P.~Uttayarat,
Phys.\ Lett.\ B {\bf 712}, 81 (2012). 

\bibitem{Cheng:2012xb}
H.Y.~Cheng and C.W.~Chiang,
Phys.\ Rev.\ D {\bf 86}, 014014 (2012). 

\bibitem{Savage:1989qr}
M.J.~Savage and R.P.~Springer,
Phys.\ Rev.\ D {\bf 42}, 1527 (1990).

\bibitem{Savage:1991wu}
M.J.~Savage,
Phys.\ Lett.\ B {\bf 257}, 414 (1991).

\bibitem{h_term}
G.~Altarelli, N.~Cabibbo and L.~Maiani,
Phys.\ Lett.\  {\bf 57B}, 277 (1975).

\bibitem{Lu:2016ogy}
C.D.~Lu, W.~Wang and F.S.~Yu,
Phys.\ Rev.\ D {\bf 93}, 056008 (2016).

\bibitem{Geng:2017esc}
C.Q.~Geng, Y.K.~Hsiao, Y.H.~Lin and L.L. Liu,
Phys.\ Lett.\ B {\bf 776}, 265 (2017). 

\bibitem{Geng:2018plk}
C.Q.~Geng, Y.K.~Hsiao, C.W.~Liu and T.H.~Tsai,
Phys.\ Rev.\ D {\bf 97}, 073006 (2018).

\bibitem{Geng:2017mxn}
C.Q.~Geng, Y.K.~Hsiao, C.W.~Liu and T.H.~Tsai,
JHEP {\bf 1711}, 147 (2017). 

\bibitem{Wang:2017azm}
W.~Wang, Z.P.~Xing and J.~Xu,
Eur.\ Phys.\ J.\ C {\bf 77}, 800 (2017). 

\bibitem{Wang:2017gxe}
D.~Wang, P.F.~Guo, W.H.~Long and F.S.~Yu,
JHEP {\bf 1803}, 066 (2018).




\bibitem{pdg}
C.~Patrignani {\it et al.} [Particle Data Group],
Chin.\ Phys.\ C {\bf 40}, 100001 (2016).

\bibitem{Brod:2011re}
J.~Brod, A.L.~Kagan and J.~Zupan,
Phys.\ Rev.\ D {\bf 86}, 014023 (2012). 

\bibitem{Brod:2012ud}
J.~Brod, Y.~Grossman, A.L.~Kagan and J.~Zupan,
JHEP {\bf 1210}, 161 (2012). 

\bibitem{Aaij:2016cfh}
R.~Aaij {\it et al.} [LHCb Collaboration],
Phys.\ Rev.\ Lett.\  {\bf 116}, 191601 (2016). 

\bibitem{Ablikim:2017ors}
M.~Ablikim {\it et al.} [BESIII Collaboration],
Phys.\ Rev.\ D {\bf 95}, 111102 (2017). 

\bibitem{ppi0}
Private communications with Dr. Peilian Li from the BESIII Collaboration
for the original data.

\bibitem{Buras:1998raa} 
A.J.~Buras, hep-ph/9806471.


\bibitem{Li:2012cfa}
H.n.~Li, C.D.~Lu and F.S.~Yu,
Phys.\ Rev.\ D {\bf 86}, 036012 (2012).

\bibitem{Fajfer:2002gp}
S.~Fajfer, P.~Singer and J.~Zupan,
Eur.\ Phys.\ J.\ C {\bf 27}, 201 (2003).

\bibitem{Abbott:1979fw}
L.F.~Abbott, P.~Sikivie and M.B.~Wise,
Phys.\ Rev.\ D {\bf 21}, 768 (1980).

\bibitem{Ablikim:2018bir}
M.~Ablikim {\it et al.} [BESIII Collaboration],
arXiv:1803.04299 [hep-ex].


\end{thebibliography}
  \end{document}